\documentclass[a4paper]{amsart}
\usepackage{graphicx}
\newcommand{\ii}{\mathrm{i}}
\newcommand{\na}{\nabla}
\newcommand{\dd}{\mathrm{d}}
\newcommand{\pd}{\partial}
\newcommand{\hh}{\mathcal{H}}
\newcommand{\U}{\mathcal{U}}
\newcommand{\M}{\mathcal{M}}

\newcommand{\e}{\mathrm{e}}
\newcommand{\ket}[1]{\left|#1\right\rangle}
\newcommand{\bra}[1]{\left\langle #1\right|}

\newcommand{\Hom}{\mathop{\mathrm{Hom}}}
\newcommand{\Iso}{\mathop{\mathrm{Iso}}}
\newcommand{\End}{\mathop{\mathrm{End}}_{\A}}
\newcommand{\tr}{\mathop{\mathrm{tr}}}

\newcommand{\Pf}{\mathop{\mathrm{Pf}}}
\renewcommand{\b}[1]{\vec{#1}}
\newcommand{\constant}{\mathrm{constant}}
\newcommand{\A}{\mathfrak{A}}
\newcommand{\B}{\mathfrak{B}}
\newcommand{\pr}[1]{\langle #1 \rangle_{\A}}
\newcommand{\prB}[1]{\langle #1 \rangle_{\B}}
\begin{document}

\title[Without Tachyon]{Exercising in K-theory: Brane Condensation without Tachyon}%
\author{Corneliu Sochichiu}%
\address{Institutul de Fizic\u a Aplicat\u a
A\c S, str. Academiei, nr. 5, Chi\c sin\u au, MD2028
MOLDOVA}%
\address{Bogoliubov Laboratory of Theoretical Physics\\
Joint Institute for Nuclear Research\\ 141980 Dubna, Moscow Reg.\\
RUSSIA}
\email{sochichi@thsun1.jinr.ru}%

\thanks{Work supported by RFBR grant \# 99-01-00190, and Scientific School support
grant}%

\begin{abstract}
We show that the $p$-dimensional noncommutative Yang--Mills model corresponding to
a $(p-1)$-brane allows solutions which correspond to lower branes. This may be
interpreted as the Morita equivalence of noncommutative planes of various
dimensions.
\end{abstract}
\maketitle
\section{Introduction}

Quantum field theory on noncommutative spaces \cite{Connes:2000by}, proved to be
an useful and powerful tool in the study of nonperturbative
strings~\cite{Seiberg:1999vs}.

Thus, the dynamics of branes in the presence of nonzero antisymmetric field
$B_{\mu\nu}$ is described by noncommutative gauge models. In the limit of large
$B_{\mu\nu}$ the noncommutativity parameter $\theta^{\mu\nu}$, is given by the
inverse matrix, $\theta^{\mu\nu}=B^{-1}_{\mu\nu}$.

In this approach the brane and string degrees of freedom are expressed in terms of
the respective noncommutative model. A definite progress was made in understanding
these models, in special, their classical solutions related to the brane
condensation (see e.g.
\cite{Gopakumar:2000zd,Sochichiu:2000rm,%
Gopakumar:2000rw,Aganagic:2000xx}).

These models exhibit new and unexpected properties in comparison with their
commutative counterparts. This some people refer to as \emph{the magic of
noncommutativity}. The noncommutative solitonic solutions, which possess no
analogues in the commutative world have the interpretation in terms of
condensation of unstable D$p$-branes to lower dimensional ones by collapsing of
certain extensions of the unstable brane.

Another manifestation of this magic is the Morita equivalence which is an
equivalence relation between different noncommutative spaces. It is believed that
Morita equivalent spaces correspond to physically equivalent situations
\cite{Connes:1998cr,Horava:1998jy,Schwarz:1998qj,Diaconescu:2000wz,%
Diaconescu:2000wy,Witten:2000cn}.

In earlier papers \cite{Sochichiu:2000ud,Sochichiu:2000bg} it was proposed that a
$p$-dimensional noncommutative Yang--Mills model with scalars can manifest itself
as a Yang--Mills model with scalar fields in a different dimension. This phenomenon
allows one to claim some equivalence relations between some noncommutative gauge
models in various dimensions. In the actual work we further elaborate on this
equivalence relation and claim that this relation can be interpreted in some sense
as a Morita equivalence. Also we propose a demonstration of this equivalence,
which in our opinion is a more natural than one proposed by the author of
\cite{Sochichiu:2000bg}, since it does not require additional alteration of the
noncommutative plane.

The solutions realising this equivalence can also be interpreted in terms of brane
condensation. As in the case of tachyon mediated brane condensation the number of
``degrees of freedom'' of collapsed and non-collapsed brane is the same, however,
in contrast to this in our case the respective dimensions collapse to zero size.

The plan of the paper is as follows. In the next section we give an alternative
construction for the solution relating $p$-dimensional noncommutative Yang--Mills
model with $d$ scalar fields to $p=2$-dimensional Yang--Mills model with $d+p-2$
scalar fields. (The total number of fields $D=p+d$ is kept fixed.) In the third
section we build the same correspondence in terms of K-theory. Finally we discuss
the results.
\section{The Equivalence}

In this section we find a solution in the $p$-dimensional noncommutative
Yang--Mills model with scalar fields, which corresponds to a two-dimensional
Yang--Mills model with scalar fields. We conventionally call this model the
``Yang--Mills--Higgs model'' and hope that there will be no confusion regarding
this notion.

Consider the model of $p$-dimensional noncommutative U(1) Yang--Mills field
interacting with $d$ real (Hermitian) scalar fields $\phi_i$ $i=1,\dots,d$, and
living on noncommutative space given by the algebra
\begin{equation}\label{noncom}
  [x^\mu,x^\nu]=\ii \theta^{\mu\nu},
\end{equation}
where we assume that the antisymmetric matrix $\theta^{\mu\nu}$ is invertible.

The model is described by the action,
\begin{equation}\label{action}
  S=\int \dd^p x \left(-\frac{1}{4g^2}F_{\mu\nu}^2-\frac{1}{2}(\na_\mu\phi_i)^2+
  \frac{1}{4g^2}[\phi_i,\phi_j]^2 \right),
\end{equation}
where,
\begin{equation}\label{f}
  F_{\mu\nu}=\pd_\mu A_\nu-\pd_\nu A_\mu+ \ii(A_\mu*A_\nu-A_\nu*A_\nu).
\end{equation}


The star product in eqs. (\ref{f}) is defined as follows,
\begin{equation}\label{star}
  (A*B)(x)=\left.\e^{-\frac{\ii}{2}
  \theta^{\mu\nu}\pd_\mu\pd_\nu'}A(x)B(x')\right|_{x'=x},
\end{equation}
$\pd_\mu$ and $\pd_\mu'$ denote derivatives with respect to
$x^\mu$ and $x^{\prime \mu}$.

The functions on noncommutative space, subject to star product (\ref{star}) realise
the representation of the Heisenberg algebra (\ref{noncom}) in terms of Weyl
ordered symbols.

One can, however, come back to the operator form. In this case the action
(\ref{action}) is rewritten in the form as follows,
\begin{equation}\label{action_op}
  S= (2\pi)^{\frac{p}{2}}|\Pf \theta|\frac{1}{4g^2}\tr (
  [X_M,X_N]^2-B^2)=\frac{1}{4\tilde{g}^2}
  \tr([X_M,X_N]^2-B^2),
\end{equation}
where capital roman indices $M,N$ span the range $1,\dots,D=p+d$. $\Pf\theta$
stands for the Pfaffian of the matrix $\theta^{\mu\nu}$,
$B^2=(\theta^{-1}_{\mu\nu})^2$, $X_\mu=\theta^{-1}_{\mu\nu}x^\nu+A_\mu$ and
$X_i=\phi_i$ are Hermitian operators acting on the Hilbert space $\hh$ on which
the Heisenberg algebra (\ref{noncom}) is represented. With the background invariant
coupling $\tilde{g}=g/\sqrt{(2\pi)^{p/2}|\Pf\theta|}$ this corresponds to the
bosonic part of the IKKT matrix model \cite{Ishibashi:1996xs}, at $N=\infty$.

Another advantage of the form (\ref{action_op}) of the action is that it is
written in the background independent form \cite{Seiberg:2000zk}, and
Yang--Mills--Higgs system with the same number of fields look similar in different
dimensions. In what follows we are going to show that noncommutative
Yang--Mills--Higgs models in different dimensions are just perturbative sectors
related to different backgrounds of the same model given by the action
(\ref{action_op}).

As usual, by a proper Lorentz transformation one can bring the ``tensor''
$\theta^{\mu\nu}$ to the canonical block-diagonal form with $i$-th,
$i=1,\dots,p/2$, $2\times 2$ antisymmetric block having $\theta_{(i)}$ as its
entry. In this case the set of operators $x^\mu$ is split in momentum and position
operators $p_i$ and $q^i$ satisfying the usual Heisenberg commutation relations
\begin{equation}
  [p_i,q^j]=-\ii \theta_{(i)}\delta_i^j,\qquad \theta_{(i)}>0 .
\end{equation}

Further, one can pass to  ``complex coordinates'' $a_i$,
$\bar{a}_i$, which are given by oscillator lowering and rising
operators,
\begin{gather}\label{abara1}
   a_i=\frac{1}{\sqrt{2\theta_{(i)}}}(q^i+\ii p_i),\qquad
   \bar{a}_i=\frac{1}{\sqrt{2\theta_{(i)}}}(q^i-\ii p_i),\\ \label{abara2}
   [a_i,\bar{a}_j]=\delta_{ij}, \qquad N_i=\bar{a}_ia_i,
\end{gather}
where in the last equation no sum is assumed. Eigenvalues of $N_i$
form an $p/2$-dimensional half-infinite lattice,
\begin{equation}\label{eigen}
  N_i\ket{\b{n}}=n_i\ket{\b{n}}, \qquad \b{n}\in\mathbb{Z}_+^{\frac{p}{2}},
\end{equation}
where $\bar{a}_i$ and $a_i$ act as rising and lowering operators
for the value $n_i$.

Equations of motion corresponding to the action (\ref{action_op})
look as follows,
\begin{equation}\label{eq_mo}
  [X_M,[X^M,X^N]]=0.
\end{equation}
In this viewpoint $X_M$ are just a set of $p$ operators acting on the Hilbert
space $\hh$ with basis formed by vectors $\ket{\b n}$. In fact an arbitrary
solution to the eq. (\ref{eq_mo}) can be interpreted either as (flat) covariant
derivatives of some noncommutative space or a constant curvature field
configuration in $p$ dimensions.

As an example consider the case $p=4$ and $d=0$. The ``complex coordinates'' are
$a_i$ and $\bar{a}_i$, $i,j=1,2$. In fact, one can consider $d\neq 0$, but the
four-dimensional scalar fields do not play any important r\^{o}le in the analysis
at this stage.

The equation (\ref{eq_mo}) has a solution,
\begin{align}\label{sol}
  &X_1=\sqrt{\frac{\theta}{2}}(A+\ii\bar{A}),\qquad
  X_2=\sqrt{\frac{\theta}{2}}(A-\ii\bar{A}) \\
  &X_3=X_4=\constant,
\end{align}
where the operators $A$ and $\bar{A}$ and the parameter $\theta$
are defined as follows.

\begin{figure}
   \includegraphics[width=80mm]{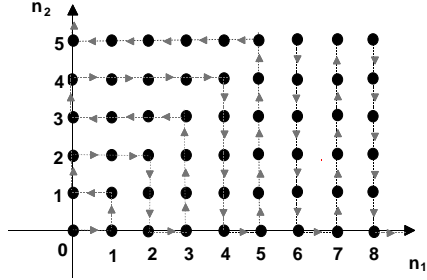}
   \caption{An example of the two-dimensional solution in the
   four-dimensional theory.
   The orbit of the operator $\bar{A}$ is shown by arrows, while $A$
   act counter the arrows. Operators $a_i$ and $\bar{a}_i$, $i=1,2$ act
   along respective axes. This is only one of
   an infinite number of possibilities to enumerate sites of the
   lattice $(n_1,n_2)$.}
   \label{fig}
\end{figure}

Consider e.g. an oriented zigzag line starting from the origin and filling the
two-dimensional quarter-infinite lattice, like one depicted in Fig.\ref{fig}.
Relabel the lattice point and the respective eigenvector by the integer value of
the length of the zigzag line from the origin to this point. That is,
\begin{align}\label{relab}
  &n\mapsto \b{n} \\
  &\ket{\b{n}}\mapsto \ket{n}=\ket{\b{n}(n)}.
\end{align}

Since the zigzag passes each point no more (and no less) than once
therefore this relabels the eigenvectors in a unique way.
Operators $A$ and $\bar{A}$ are defined by,
\begin{align}\label{AA1}
  & A \ket{n}=\sqrt{n}\ket{n-1},\\ \label{AA2}
  &\bar{A}\ket{n}=\sqrt{n+1}\ket{n+1},\\
  &[A,\bar{A}]=1,
\end{align}
where corresponding to the Figure \ref{fig}, $A$ is moving towards the origin
(against the arrows on the picture) and $\bar{A}$ moving from origin (along the
arrows). Operators $A$ and $\bar{A}$ can be expressed as functions on (normal
symbols of) $a_i$ and $\bar{a}_i$. We do not know the exact analytic expressions
for this functions corresponding to the case depicted in Fig.\ref{fig},
fortunately we do not need it.

Requiring finiteness of the action (\ref{action_op}) computed on the solution
(\ref{AA1}), (\ref{AA2}) one has for the noncommutativity parameter,
$\theta=\sqrt{2/(B^2)}$.

Having solution  (\ref{relab}-\ref{AA2}) one can introduce the Weyl ordered symbol
with respect to $A$ and $\bar{A}$ for some well behaved operator $\Phi$, e.g. one
satisfying $\tr \Phi^* \Phi<\infty$. The respective Weyl symbol $\Phi(A,\bar{A})$
defines a function on the \emph{two-dimensional} noncommutative plane,
\begin{align}\label{2dim}
  &\Phi(A,\bar{A})=\frac{1}{2\pi}\int \dd^2k\ \e^{ 2\ii(\bar{A}k+\bar{k}A)}
  \tr (\e^{-2\ii(\bar{A}k+\bar{k}A)}\Phi), \\ \label{2dim2}
  &(\Phi*\Psi)(A,\bar{A})=\e^{-\frac{1}{2}(\pd\bar{\pd}'-\bar{\pd}\pd')}
  \Phi(A,\bar{A})\Psi(A',\bar{A}')\left|_{A'=A \atop \bar{A}'=\bar{A}}\right.,
\end{align}
where the integration in (\ref{2dim}) is performed through the
two-dimensional (commutative) complex plane $(k,\bar{k})$.

This gives formulas for passing from functions on the
four-dimensional noncommutative plane to functions on the
two-dimensional one. By construction this procedure is invertible
and therefore it establishes an equivalence relation two algebras
of noncommutative functions \cite{Sochichiu:2000bg}. In next
section we are going to show that this is in fact a Morita type
equivalence.

The solution (\ref{relab}-\ref{AA2}) is at no way unique. All possible solutions
of this type are parameterised by different ways in choosing ordered basis in the
Hilbert space.

In the case of arbitrary even $p\leq D$ one obtains such a solution by enumerating
the basis of the separable Hilbert space and defining $A$ and $\bar{A}$ according
to (\ref{AA1}) and (\ref{AA2}) where $n$ is the number of the basis element.

The arbitrariness in choosing an orthonormal basis in the Hilbert space $\hh$ is
parameterised by the unitary operator $U\in \U(\hh)$, where $\U(\hh)$ is the
unitary group of the Hilbert space $\hh$, sometimes called U$(\infty)$. Therefore
$\U(\hh)$ also is the moduli space of the map (\ref{relab}-\ref{AA2}). As it is
clear this dependence however can be eaten by a gauge transformation of either
two-dimensional or $p$-dimensional model.

So far we the transformation of scalar noncommutative functions. For a non-scalar
function e.g. a vector one but other than the gauge field one has, roughly
speaking, two two-dimensional vector components and $(p-2)$ scalars. The problem
is how to split it in the vector and scalar components. A priori, there is no
restriction to do this, the total arbitrariness being by the the Grassman manifold
SO(p)/SO(2)$\times$SO(p-2).

Having this one may be tempted to translate from $p$ to two dimensions also the
small fluctuations of the gauge fields $A_\mu$. In trying to do this there is a
problem. By the above construction one can translate only quantities which
transform covariantly under the action of the noncommutative U(1). (i.e. such
quantities which can be represented by a background independent operator.) It is
known that $A_\mu$ does not transform covariantly. However, the quantity
$X_\mu=p_\mu+A_\mu$, $\mu=1,\dots,p$ do and can be written in the two dimensional
form. But when one will try to go back from the description in terms of $X$'s to
the description in terms of two dimensional gauge fields and scalars one will
realise that neither of the fields do vanish at the infinite, although the action
is finite. This in fact means that there are different perturbative regimes
corresponding to solutions giving different dimensionalities.

Let us note, that in order to keep the action (\ref{action}) invariant under this
redefinition one has to require,
\begin{equation}\label{inv_ac}
   (2\pi)^{\frac{p}{2}}\Pf \theta_{(p)}\frac{1}{4g_{(p)}^2}=
   2\pi \theta_{(2)}\frac{1}{4g_{(2)}^2},
\end{equation}
where the subscript in the parentheses denotes the dimension to which the quantity
refers. This condition gives for the two dimensional gauge coupling,
\begin{equation}
   g_{(2)}^2=\frac{\sqrt{2}}{(2\pi)^{\frac{p}{2}-1}B\Pf \theta_{(p)}}g_{(p)}^2.
\end{equation}

The arguments above can be turned back, i.e. one can consider the initial $p$
dimensional noncommutative space (\ref{noncom}) as a solution in the two
dimensional model defined by (\ref{sol}) and (\ref{AA1},\ref{AA2}). This was
originally the way proceeded in ref \cite{Sochichiu:2000bg}.

We came thus to the natural conclusion that noncommutative Yang--Mills--Higgs
models in different dimensions behave like different perturbative regions of the
same background independent model (\ref{action_op}) \cite{Seiberg:2000zk}, defined
in terms of operators acting on some abstract separable Hilbert space $\hh$.
\section{K-theory meaning}

K-theory toolkit seems to be appropriate for the study of brane dynamics as well
as of noncommutative gauge models
\cite{Horava:1998jy,Schwarz:1998qj,Diaconescu:2000wz,Diaconescu:2000wy,Witten:2000cn}.
In this section we are going to exploit some K-theory tools in order to understand
the results of the previous section from this point of view. In order to use them
let us make following
\cite{Horava:1998jy,Schwarz:1998qj,Diaconescu:2000wz,Diaconescu:2000wy,Witten:2000cn}
a very short review of K-theory in application to the noncommutative geometry.

Consider an associative complex algebra $\A$ with involution ``$^*$'' (a
$C^*$-algebra). We will mainly think about the algebra of complex functions on the
noncommutative plane. In our case it is the Heisenberg algebra. Let $E$ be its
left module i.e.,
\begin{equation}\label{left_m}
  a(m)=am\in E,\qquad (a'a)(m)=a'(am)=a'am,
\end{equation}
for arbitrary $m\in E$, and $a,a'\in \A$. Right module is defined
in a similar way but with consequent action of elements of $\A$
from the right.

The algebra $\A$ itself as well as $n$ copies of it $\A\oplus\A\oplus\dots\oplus\A$
is a primitive example of both left and right modules, such modules are called
\emph{free}. A module $E$ for which exists another module $E'$ such that $E\oplus
E'$ is free is called a \emph{projective} one. (It is clear that $E'$ is also a
projective module.) The set of left or right projective modules form a semigroup
with respect to the direct sum operation. This semigroup can be ``upgraded'' to a
group as follows.

Consider pairs of modules $(E,F)$, with the composition rule
$(E,F)+(E',F')=(E\oplus E',F'\oplus F)$ and the equivalence
relation $(E,F)\sim (E\oplus G,F\oplus G)$, for arbitrary module
$G$. This equivalence classes already form a group whose unity is
given by $(G,G)$-pairs and the opposite element to $(E,F)$ given
by $(F,E)$,
\begin{equation*}
  (E,F)+(F,E)=(E\oplus F,E\oplus F)\sim (G,G).
\end{equation*}
This trick is similar to one used to extend the set of positive numbers to real
ones. The group one gets in a such way is called the K$(\A)$, or, if $\A$ is the
algebra of functions on some space $\M$ it is denoted as K$(\M)$.

Let us equip our left or right projective module $E$, with an
$\A$-valued product $\pr{\ ,}$, satisfying,
\begin{align}\label{prod1}
  &\pr{m,m'}^*=\pr{m',m} \\ \label{prod2}
  &\pr{am,m'}=a\pr{m,m'} \\ \label{prod3}
  &\pr{m,m'}\text{ is a positive element in }\A.
\end{align}
The $\A$-module $E$ is called \emph{full} when the linear span of the range of
$\pr{\ ,}$ is dense in $\A$.

One can introduce connection $\na_\alpha$ on the $E$ with respect to some element
of the algebra of infinitesimal automorphisms of $\A$: $a\to a+\delta_\alpha a$,
labelled by some element $\alpha$, which satisfies,
\begin{equation}\label{connection}
  \na_\alpha(am)=a\na_\alpha(m)+(\delta_\alpha a) m,
\end{equation}
and it is linear in $\alpha$. Using this connection one can built
the curvature associated to it,
\begin{equation}\label{curvature}
  F_{\alpha\beta}=[\na_\alpha,\na_\beta]-\na_{[\alpha,\beta]}.
\end{equation}

$\A$-linear maps $T:E\to E$ which have an adjoint with respect to
the product (\ref{prod1}-\ref{prod3}) and commute with the action
of $\A$ on $E$ form the algebra $\End E$ of endomorphisms of the
$\A$-module $E$.

By definition an algebra $\B$ is Morita equivalent to $\A$ if it
is isomorphic to $\End E$ for some complete module $E$.

There exists the following criterium for Morita equivalence of two algebras $\A$
and $\B$. A left $\A$-module $P$ which is also a right $\B$-module is called
$(\A,\B)$-bimodule. Assume that $P$ as $\A$- and $\B$-module is equipped with
$\A$-valued product $\pr{\ ,}$, and $\B$-valued product $\prB{\ ,}$, and it is
full as both $\A$- and $\B$-module. When it exists such a module is called
$(\A,\B)$ \emph{equivalence bimodule}, in this case algebras $\A$ and $\B$ are
Morita equivalent. The Morita equivalence allows one to establish relations
between various structures of the equivalent algebras and their modules, like
endomorphisms, connections, etc.

It is conjectured
\cite{Horava:1998jy,Schwarz:1998qj,Diaconescu:2000wz,Diaconescu:2000wy,%
Witten:2000cn}, that Morita equivalent algebras in string theory correspond to
physically equivalent systems e.g. related by duality transformations. In
noncommutative theory the gauge models on the dual tori are also known to be
Morita equivalent \cite{Connes:1998cr,Schwarz:1998qj}.

Let us return back to the model (\ref{action_op}). The algebra $\A_p$ now is one
generated $x^\mu$ subject to commutation relation (\ref{noncom}), or in
alternative basis, respectively, by $p/2$-dimensional oscillator rising and
lowering operators $\bar{a}_i,a_i$ (\ref{abara1},\ref{abara2}). We will use the
last choice. In this case the $p/2$-dimensional oscillator Hilbert space $\hh_p$
with the basis $\ket{\b{n}}$ plays the role of a complete $\A$-module.

Consider the space $P=\hh_2\otimes\hh^*_p$  which is the linear span of elements
$\ket{n}\bra{\b{n}}$, where $\ket{\b{n}}\in\hh_{2}$ and $\bra{\b{n}}\in
\hh_{p}^*$. $P$ is at the same time left module for one-dimensional oscillator
algebra (two-dimensional noncommutative functions) and right module for the $p/2$
dimensional oscillator algebra ($p$-dimensional noncommutative plane function
algebra). As both $\A_p$ and $\A_2$ module $P$ is complete, hence it is an
equivalence one. Therefore, the function algebra of the $p$-dimensional
noncommutative plane is in some sense\footnote{It seems that the Hilbert space
itself is not finitely generated projective module.} Morita equivalent to one on
the two-dimensional noncommutative plane.

In the case of Morita equivalent algebras one has a correspondence between the
$\A_p$ and $\A_2$ modules. Thus for an $\A_p$-module $\hh_p$ one has an
$\A_2$-module $\hh_2$ given by,
\begin{equation}\label{d->2}
  \hh_2=P\otimes_{\A_p}\hh_p,
\end{equation}
where the tensor product with respect to $\A_p$ is obtained from usual (complex)
tensor product $\otimes$ by means of identification, $pa\otimes m\sim p\otimes
am$, $a\in\A_p$.

It seems to be a problem because neither of modules $\hh_p$ or $P$ seems to be
\emph{finitely generated projective}. However, one can avoid this problem by
considering a \emph{regularised} system. One can regularise the algebras
(\ref{abara1},\ref{abara2}) and (\ref{AA1},\ref{AA2}), e.g. by $q$-deforming them
as proposed in \cite{Donets:2000ic} with $q^N=1$,
\begin{equation}\label{reg1}
  a_i\ket{\b{n}}=\sqrt{\frac{1}{\pi}\sin\frac{\pi n_i}{N}}
  \ket{\b{n}-\b e_i},\qquad
  \bar{a}_i\ket{\b{n}}=\sqrt{\frac{1}{\pi}\sin\frac{\pi (n_i+1)}
  {N}}\ket{\b{n}+\b e_i},
\end{equation}
where $\b e_i$ is the $i$-th unit lattice vector. Then the limit $N\to\infty$
corresponds to the ``cut-off'' removing. In this case the regularised Hilbert
space and the equivalence module become finite dimensional and finitely generated
projective. Moreover, irrelevant to $N$ all regularised models fall in the same
Morita equivalence class and one can define the cut-off removed model as an
extremal element of the extremal element of this class.

However, as noted in \cite{Witten:2000cn}, there is a difference
between finite dimensional cases and the case $N=\infty$. In
particular all infinite dimensional separable Hilbert spaces are
known to be isomorphic. In this case establishing Morita
equivalence  is equivalent to establishing all possible maps
between the $\A$ and $\B$ modules.

Indeed, due to the irreducibility of the action of the Heisenberg algebra $\A_{p}$
on the Hilbert space, the equivalence module $P$ can be represented as a tensor
product $\hh_{2}\otimes \hh^*_p$ which is isomorphic to $\Hom (\hh_p,\hh_2)$. This
set contains all maps from the Hilbert space of the $p$-dimensional Heisenberg
algebra to the Hilbert space of the 2-dimensional one. In the previous section we
considered only isomorphic maps: $\Iso (\hh_p,\hh_2)\subset \Hom (\hh_p,\hh_2)$.
Since it is Hilbert spaces which are mapped, the set of isomorphic maps preserving
the Hilbert space product is in its turn isomorphic to the infinite dimensional
unitary group: $\Iso (\hh_p,\hh_2)\cong \U(\hh_p)\cong\U(\hh_2)$.

This gives exactly the moduli of the equivalence of $p$- and 2-dimensional models
described in the previous section.
\section{Discussions and Conclusions}

In this paper we have shown that a $p$-dimensional noncommutative gauge model with
scalar fields possesses solutions which can be interpreted as a 2-dimensional gauge
model. We considered nondegenerate solutions, i.e. ones which realise a isomorphism
(equivalence) between two models. The moduli of such solutions are given by the
group of unitary transformations of the separable Hilbert space. This means that
the isomorphism is unique up to a noncommutative U(1) gauge transformation.

The construction of the second section can be turned back, i.e. one can consider
solutions in the 2-dimensional gauge model with scalars which behaves like a
$p>2$-dimensional model with less scalars. This suggests that the equivalence is
valid for all noncommutative models of the types given by the action
(\ref{action}) in even dimensions less or equal $D$, which is the total number of
fields and with the same factor $B^2=(\theta_{\mu\nu}^{-1})^2$.

The K-theory considerations allow one to generalise the above solutions and to
consider arbitrary maps from the $p$-dimensional model to another $p'$-dimensional
one and back while both $p$ and $p'$ are even. These include also the projector
solutions, i.e. ones when the orbit of $A$ and $\bar{A}$  span only a sublattice
of the $p/2$-dimensional lattice of the eigenvalues of $N_i$. Such solutions should
have some relevance to the noncommutative solitons
\cite{Gopakumar:2000zd,Sochichiu:2000rm,Aganagic:2000xx,Harvey:2000jb}.

In the string theory picture the gauge fields present in the model we considered
describe the tangential coordinates of a $(p-1)$-brane while the scalar fields
correspond to the transversal ones. The interpretation of the obtained solutions,
therefore is as a brane whose number of extensions collapsed to zero size or
oppositely as some new dimensions have been blown up. This is similar to the
tachyon condensation picture, however, there is no tachyonic mode in the spectrum
of the model since we are considering BPS and therefore stable solutions. The
other difference is that in our case the the thickness of a condensed brane is
exactly zero and not finite one as in the tachyon condensation case. Also, the
tachyon condensation mechanism cannot provide appearance of new extensions to a
brane in contrast to our case.

Due to the described equivalence we have a single operator model described by the
action (\ref{action_op}) with an abstract separable Hilbert space rather than
noncommutative gauge models in different dimensions. The respective gauge models
are given by fluctuations around particular solutions in the main operator model.

One may be surprised by the fact that the gauge models have different
renormalisation behaviour in different dimensions. In particular one has different
external divergence indices for the same Feynmann diagrams in different
dimensions.

In fact there is no contradiction if to observe that the respective noncommutative
models correspond to different points of perturbative expansion of the main
operator model. It is usual that the expansion around different points may have
different convergence properties, but due to the IR/UV mixing one may presume that
nonperturbatively, or even up to all orders in perturbation theory the model has
the same renormalisation behaviour in all dimensions.


\begin{thebibliography}{10}

\bibitem{Connes:2000by}
A.~Connes, {\it A short survey of noncommutative geometry},  {\em J. Math.
  Phys.} {\bf 41} (2000) 3832--3866,
  [\href{http://xxx.lanl.gov/abs/hep-th/0003006}{{\tt hep-th/0003006}}].

\bibitem{Seiberg:1999vs}
N.~Seiberg and E.~Witten, {\it String theory and noncommutative geometry},
  {\em JHEP} {\bf 09} (1999) 032,
  [\href{http://xxx.lanl.gov/abs/hep-th/9908142}{{\tt hep-th/9908142}}].

\bibitem{Gopakumar:2000zd}
R.~Gopakumar, S.~Minwalla, and A.~Strominger, {\it Noncommutative solitons},
  {\em JHEP} {\bf 05} (2000) 020,
  [\href{http://xxx.lanl.gov/abs/hep-th/0003160}{{\tt hep-th/0003160}}].

\bibitem{Sochichiu:2000rm}
C.~Sochichiu, {\it Noncommutative tachyonic solitons: Interaction with gauge
  field},  {\em JHEP} {\bf 08} (2000) 026,
  [\href{http://xxx.lanl.gov/abs/hep-th/0007217}{{\tt hep-th/0007217}}].

\bibitem{Gopakumar:2000rw}
R.~Gopakumar, S.~Minwalla, and A.~Strominger, {\it Symmetry restoration and
  tachyon condensation in open string theory},
  \href{http://xxx.lanl.gov/abs/hep-th/0007226}{{\tt hep-th/0007226}}.

\bibitem{Aganagic:2000xx}
M.~Aganagic, R.~Gopakumar, S.~Minwalla, and A.~Strominger, {\it Unstable
  solitons in noncommutative gauge theory},
  \href{http://xxx.lanl.gov/abs/hep-th/0009142}{{\tt hep-th/0009142}}.

\bibitem{Connes:1998cr}
A.~Connes, M.~R. Douglas, and A.~Schwarz, {\it Noncommutative geometry and
  matrix theory: Compactification on tori},  {\em JHEP} {\bf 02} (1998) 003,
  [\href{http://xxx.lanl.gov/abs/hep-th/9711162}{{\tt hep-th/9711162}}].

\bibitem{Horava:1998jy}
P.~Horava, {\it Type {IIA D-branes, K-theory}, and matrix theory},  {\em Adv.
  Theor. Math. Phys.} {\bf 2} (1999) 1373,
  [\href{http://xxx.lanl.gov/abs/hep-th/9812135}{{\tt hep-th/9812135}}].

\bibitem{Schwarz:1998qj}
A.~Schwarz, {\it Morita equivalence and duality},  {\em Nucl. Phys.} {\bf B534}
  (1998) 720--738, [\href{http://xxx.lanl.gov/abs/hep-th/9805034}{{\tt
  hep-th/9805034}}].

\bibitem{Diaconescu:2000wz}
D.-E. Diaconescu, G.~Moore, and E.~Witten, {\it A derivation of {K-theory} from
  {M-theory}},  \href{http://xxx.lanl.gov/abs/hep-th/0005091}{{\tt
  hep-th/0005091}}.

\bibitem{Diaconescu:2000wy}
D.-E. Diaconescu, G.~Moore, and E.~Witten, {\it {E(8)} gauge theory, and a
  derivation of {K-theory} from {M-theory}},
  \href{http://xxx.lanl.gov/abs/hep-th/0005090}{{\tt hep-th/0005090}}.

\bibitem{Witten:2000cn}
E.~Witten, {\it Overview of {K-theory} applied to strings},
  \href{http://xxx.lanl.gov/abs/hep-th/0007175}{{\tt hep-th/0007175}}.

\bibitem{Sochichiu:2000ud}
C.~Sochichiu, {\it {M(any) vacua of IIB}},  {\em JHEP} {\bf 05} (2000) 026,
  [\href{http://xxx.lanl.gov/abs/hep-th/0004062}{{\tt hep-th/0004062}}].

\bibitem{Sochichiu:2000bg}
C.~Sochichiu, {\it On the equivalence of noncommutative models in various
  dimensions},  {\em JHEP} {\bf 08} (2000) 048,
  [\href{http://xxx.lanl.gov/abs/hep-th/0007127}{{\tt hep-th/0007127}}].

\bibitem{Ishibashi:1996xs}
N.~Ishibashi, H.~Kawai, Y.~Kitazawa, and A.~Tsuchiya, {\it A {large-N} reduced
  model as superstring},  {\em Nucl. Phys.} {\bf B498} (1997) 467,
  [\href{http://xxx.lanl.gov/abs/hep-th/9612115}{{\tt hep-th/9612115}}].

\bibitem{Seiberg:2000zk}
N.~Seiberg, {\it A note on background independence in noncommutative gauge
  theories, matrix model and tachyon condensation},  {\em JHEP} {\bf 09} (2000)
  003, [\href{http://xxx.lanl.gov/abs/hep-th/0008013}{{\tt hep-th/0008013}}].

\bibitem{Donets:2000ic}
E.~E. Donets, A.~P. Isaev, C.~Sochichiu, and M.~Tsulaia, {\it Brane vacuum as
  chain of rotators},  {\em JHEP} {\bf 12} (2000) 022,
  [\href{http://xxx.lanl.gov/abs/hep-th/0011090}{{\tt hep-th/0011090}}].

\bibitem{Harvey:2000jb}
J.~A. Harvey, P.~Kraus, and F.~Larsen, {\it Exact noncommutative solitons},
  {\em JHEP} {\bf 12} (2000) 024,
  [\href{http://xxx.lanl.gov/abs/hep-th/0010060}{{\tt hep-th/0010060}}].

\end{thebibliography}
\providecommand{\href}[2]{#2}\begingroup\raggedright\endgroup
\end{document}